\documentclass[authoryear,12pt,revtex]{elsarticle}

\usepackage{amssymb,hyperref}

\journal{Studies in the History and Philosophy of Modern Physics}

\begin{document}

\begin{frontmatter}



\title{Quantum Gravity: Meaning and Measurement}


\author[John]{John Stachel}
\ead{john.stachel@gmail.com}
\author[Kaca]{Ka\'{c}a Bradonji\'{c}}
\ead{kaca.bradonjic@gmail.com}

\address[John]{Center for Einstein Studies, Boston University, Boston, MA 02215 USA}
\address[Kaca]{Physics Department, Wellesley College, 106 Central Street, Wellesley, MA 02481, USA}

\begin{abstract}
A discussion of the meaning of a physical concept cannot be separated from discussion of the conditions for its ideal measurement. We assert that quantization is no more than the invocation of the quantum of action in the explanation of some process or phenomenon, and does not imply an assertion of the fundamental nature of such a process. This leads to an ecumenical approach to the problem of quantization of the gravitational field. There can be many valid approaches, each of which should be judged by the domain of its applicability to various phenomena. If two approaches have overlapping domains, the relation between them then itself becomes a subject of study. We advocate an approach to general relativity based on the unimodular group, which emphasizes the physical significance and measurability of the conformal and projective structures. A discussion of the method of matched asymptotic expansions, and of the weakness of terrestrial sources compared with astrophysical and cosmological sources, leads us to suggest theoretical studies of gravitational radiation based on retrodiction (observation) rather than prediction (experimentation).
\end{abstract}


\end{frontmatter}


\section{Introduction}
\label{s:intro}
How can we combine the background-independent, dynamical approach to all space--time structures of general relativity with the quantum theory, which is based on fixed, absolute background space--time structures? That is the fundamental challenge of quantum gravity.

Most approaches to quantum gravity concentrate on the development of a formalism, and only then take up the question of physical applications of this formalism. But this division neglects one of the most important lessons that can be drawn from the history and philosophy of science. No one has stated this lesson more eloquently than Gaston Bachelard: 

\begin{quotation}
In order to embody new experimental evidence, it is necessary to deform the original concepts, study the conditions of applicability of these concepts, and above all incorporate the conditions of applicability of a concept into the very meaning of the concept. ... The classic division that separates a theory from its application ignores this necessity to incorporate the conditions of applicability into the very essence of the theory.~\citep[pg. 61; transl. by J.S.]{bachelard}
\end{quotation}

We shall discuss some aspects of the problem of quantization of the gravitational field equations in the light of the need to combine the mathematical definition of physically meaningful candidates for quantization with the description of conditions for their measurement in principle, or as we shall say, ideal measurement. Again we may draw inspiration from the words of Bachelard:

\begin{quotation}
I believe myself that mathematical thought forms the basis of physical explanation and that the conditions of abstract thought from now on are inseparable from the conditions of scientific experiment~\citep[p. 131, translation from \citep{lecourt}, p. 57]{bachelard}
\end{quotation}

What we shall present here is not a single theory, but a research program. As Imre Lakatos states in his ground-breaking paper ``Falsification and the Methodology of Scientific Research Programmes,"
\begin{quotation}
Sophisticated falsificationism thus shifts the problem of how to appraise theories to the problem of how to appraise series of theories. Not an isolated theory, but only a series of theories can be said to be scientific or unscientific: to apply then the term ``scientific" to one single theory is a category mistake. The time-honored empirical criterion for a satisfactory theory was agreement with the observed facts. Our empirical criterion for a series of theories is that it should produce new facts. The idea of growth and the concept of empirical character are soldered into one (Lakatos, 1970).
\end{quotation}

\section{What is Quantization?}
\label{quantization}

Given a physical theory, what elements of it to quantize is not an obvious question. First of all, one must decide what quantization means. One of us has emphasized that quantization consists of some procedure used to take account of the existence of $h$, the quantum of action:

\begin{quotation}
Quantization is just a way accounting for the effects of $h$, the quantum of action, on any process involving some system, or rather on theoretical models of such a system, fundamental or composite; in the latter case, the collective behavior of a set of more fundamental entities is quantized. Successful quantization of some classical formalism does not mean that one has achieved a deeper understanding of reality -- or better, an understanding of a deeper level of reality. It means that one has successfully understood the effects of the quantum of action on the phenomena (processes) described by the formalism.

The search for a method of quantizing space--time structures associated with the Einstein equations is quite distinct from the search for an underlying theory of \emph{all} ``fundamental" interactions. An attempt to quantize one set of space--time structures does not negate, and need not replace, attempts to quantize another set of space--time structures. Everything depends on the utility of the results in explaining some physical processes.\citep{stachel12}
\end{quotation}

There are many such examples of different approaches to quantization of the same physical process, each successful within its range of applicability~\citep{stachel12}. Rather than leading to a contradiction, this leads to a new and interesting question: what is the relation between such two approaches? For example, Crenshaw has shown that there is a ``limited equivalence between microscopic and macroscopic quantizations of the electromagnetic field in a dielectric"~\citep{crenshaw}. Another example is the relation between loop quantization and the usual field quantization of the electromagnetic field -- if the loops are ``thickened," the two are equivalent~\citep{ashtekar}.

\section{Measurability Analysis}
\label{manalysis}
The formal quantization procedure adopted may not be unique, and may even involve quantities, such as gauge-dependent variables, that are not measurable even in principle. But the physically significant upshot must be to single out a class of physical quantities that are measurable in principle. Following Peter Bergmann, we shall call this the problem of measurability analysis: 
\begin{quotation}
Measurability analysis identifies those dynamic field variables that are susceptible to observation and measurement (``observables"), and investigates to what extent limitations inherent in their experimental determination are consistent with the uncertainties predicted by the formal theory.~\citep{bergman}
\end{quotation}

Measurability analysis identifies those concepts that a theory defines as meaningful within some context and investigates to what extent the values associated with these concepts are ideally measurable in the defining context. It is just as applicable to classical as it is to quantum theories. For example, one can study the differing conditions of applicability of concepts such as hardness and viscosity in the context of the fluid and solid states of matter in classical thermodynamics \citep{stachel86}. One must always establish a qualitative and quantitative \emph{consonance} between the concept of some \emph{entity}, to which physical significance is ascribed, and an ideal \emph{measurement procedure} for that entity.

Indeed, it seems essential to first investigate the conditions of applicability of a concept in a classical context, if it has one, before studying any modifications in the quantum context. If it is a strictly quantum concept, $h$ must enter \emph{both} its definition \emph{and} measurement procedure. This division between classical and quantum concepts is not without its own problems. For example, it is often claimed that spin is a purely quantum concept, but it has been shown that a spin vector can be attached to a classical particle~\citep{stachel77}.

When applied to a quantum theory, measurability analysis aims to predict the effect of the quantum of action on measurements: first on individual measurements and then, perhaps even more importantly, on conjoint measurements of pairs of such quantities. This should not be confused with the so-called ``measurement problem" in quantum mechanics. We take the position that the task of any quantum theory is to predict the outcome of a process by calculating a probability amplitude for the process. The wave function is no more than a mathematical tool that is sometimes useful in such a calculation, and to which no ontological significance should be attributed. But regardless of one's opinion about either, the distinction between the measurement problem and measurability analysis should be clear.

The origin of measurability analysis for quantum mechanics can be traced back to the work of Heisenberg, and for quantum field theory to the work of Bohr and Rosenfeld. Indeed, as Bergmann and Smith emphasized, much can still be learned about the problems of quantum gravity from the Bohr--Rosenfeld (B--R) analysis of the measurability and co-measurability of the components of the electromagnetic field~\citep{bohrrosenfeld33,bohrrosenfeld78, bohrrosenfeld50, darrigol}). 
The criterion of consonance between definability and measurability of a physical quantity can play a heuristic role in the search for a viable theory of quantum gravity: 
\begin{quotation}
For well-established theories, this criterion can be tested. For example, in spite of a serious challenge, source-free quantum electrodynamics was shown to pass this test. In the case of quantum gravity, our situation is rather the opposite. In the absence of a fully accepted, rigorous theory, exploration of the limits of measurability of various quantities can serve as a tool to provide clues in the search for such a theory: If we are fairly certain of the results of our measurability analysis, the proposed theory must be fully consistent with these results.~\citep{camelia}
\end{quotation}

The first important conclusion that can be drawn from B--R is that the field components at a point are not measurable. Even an ideal measurement involves a finite region of space and takes a finite amount of time. In a word: only averages over some region of space--time are measurable and such a measurement requires a four-volume element. This is just what B--R did in their analysis of the measurability of the components of the electric and magnetic fields. 

The second important conclusion that can be drawn from (B--R) is that co-measurability of averages over two time-like-separated regions must be investigated. This suggests that formulations of a theory based on canonical commutation relations on a space-like hypersurface may not be the best starting point for an approach based on measurability analysis. Indeed, as DeWitt first emphasized~\citep{dewitt62}, the Peierls bracket provides a starting point that is much better suited to such an approach. It is only defined for quantities measurable in principle, and the definition applies to any two points in space--time. As DeWitt points out, hanging on to the canonical formulation just complicates things:
\begin{quotation}
When expounding the fundamentals of quantum field theory physicists almost universally fail to apply the lessons that relativity theory taught them early in the twentieth century. Although they usually carry out their calculations in a covariant way, in deriving their calculational rules they seem unable to wean themselves from canonical methods and Hamiltonians, which are holdovers from the nineteenth century, and are tied to the cumbersome (3+1)-dimensional baggage of conjugate momenta, bigger-than-physical Hilbert spaces and constraints. One of the unfortunate results is that physicists, over the years, have almost totally neglected the beautiful covariant replacement for the canonical Poisson bracket that Peierls invented in 1952~\citep[Introduction]{dewitt05}. 
\end{quotation}

As noted above, space--time averages must be taken to get physically valid co-measurability results. Contrary to the previous assertions of Heisenberg~\citep{heisenberg}, who only considered spatial averages at the same time, B--R showed that the average values of any pair of components of the electric and magnetic fields are co-measurable over the same region of space--time. The quantum of action only limits the co-measurability of such averages over two different space--time regions that are time-like separated. 

\section{What is physically meaningful in quantum gravity?}
\label{meaning}

Both these conclusions are quite general, and hence should apply to any theory of quantum gravity based on a differentiable manifold~\footnote{We shall not discuss here theories based on the \emph{a priori} introduction of a discrete structure for space--times.}. Now we come to some specific questions about quantum gravity. In classical general relativity, what are the physically meaningful quantities that should be quantized? Again, B--R analysis provides an important hint: they don't attempt to quantize the components of $A_{\mu}$, the non-measurable electromagnetic four-potential\footnote{Locally non-measurable; globally, of course, the two Aharonov-Bohm effects are measurable.}, which so often is the starting point of other approaches; but start immediately from $F_{\mu\nu}$, the electromagnetic field. Indeed, its components in any inertial frame of reference (ifr) are measurable as the electric and magnetic fields w.r.t. that ifr. Mathematically, $F_{\mu\nu}$ can be regarded as the curvature corresponding to the connection $A_{\mu}$. So the analogy with the B--R approach to electromagnetism suggests one start, not from the space--time connection (let alone the metric), but from the curvature tensor, or rather the physical components of the curvature tensor with respect to some orthonormal basis (see~\citep{dewitt62} and~\citep{bergman}). DeWitt has shown how to apply the Peierls bracket formalism to this case~\citep{dewitt05}. 

But another question immediately arises: \emph{which} curvature tensor? As we shall see there are two basic space--time curvature tensors:
\begin{enumerate}
\item{the conformal curvature tensor, derived from the conformal connection, and}
\item{the projective curvature tensor, derived from the projective connection.}
\end{enumerate}

We shall return to this question, but for the moment let us just note that in our view both are relevant -- the answer depends on whether we are concerned with the near field, tied to its sources, or the far field, which has escaped from its sources, i.e., the free-radiation field. 

Now we come to the final crucial question. Bohr emphasized the need to base measurability analysis on some massive physical materialization of an inertial frame of reference, to which the measuring instruments were either rigidly attached, or w.r.t. which their motion could be related~\citep{bohr35,bohr49}. In particular, the four-volume element used in B--R is defined by a charged test body, initially at rest in the ifr, and then set into rigid inertial motion by its interaction with the field for a finite time interval. The entire analysis is based on the existence of a fixed Minkowski space--time, and the fact that the presence of massive bodies -- charged like the test body, or uncharged like the massive body defining the ifr -- have no effect on that space--time. In other words, electromagnetic theory is a background-dependent theory. In addition, the ability to make the inertial mass of a body as large as needed while keeping its charge fixed, plays an important role in the B--R analysis, enabling them to minimize the effect of radiation reaction on the test body. 
	
General relativity differs in both respects: there are no fixed, non-dynamical structures on the differentiable manifold\footnote{Even its global topology is not fixed.}, and the presence of any massive body, charged or uncharged, and even of the electromagnetic field itself, has an effect on the space--time structure prescribed by the stress--energy--momentum tensor through the inhomogeneous Einstein equations. In other words, general relativity is a background-independent theory. In addition, as first noted by \citet{bronstein}, since gravitational charge and inertial mass are proportional (the equivalence principle), if one makes the inertial mass too large, the body will fall within its own Schwarzschild radius and become unavailable as a test body.

 \section{How to quantize in general relativity?}
How can one proceed to quantize such a theory? We can get a first hint by noting that the above argument of Bronstein is only an argument against attempting to measure the components of the gravitational portion of the affine connection w.r.t. a global frame of reference, which divides the connection into an inertial connection and a gravitational tensor. The argument disappears once we move to the level of the curvature tensor, because 
there is a big difference between the measurability of $F_{\mu\nu}$, the electromagnetic field tensor, and of $R^{\kappa}_{\lambda\mu\nu}$, the gravitational curvature tensor. As noted above, to measure the components of $F_{\mu\nu}$, the electric and magnetic fields, we need to fix an inertial frame of reference and measure motions of test bodies w.r.t. it. The components of $R^{\kappa}_{\lambda\mu\nu}$, on the other hand, relate to tidal gravitational forces, measurement of which does not require an overall global frame of reference. We only need to measure the relative acceleration of two test bodies with respect to each other when both are in free, inertio-gravitational motion, or even the relative acceleration of the parts of one test body if its stress--energy--momentum tensor is known.
 
And now the equivalence principle turns into an advantage: since passive inertial and active gravitational masses are equal, the motion of a test body used to probe this relative acceleration field is independent of its passive mass; so the mass of the test body may be made sufficiently small that the effect of its active gravitational mass on the field being investigated is either negligible; or if its perturbing effect cannot be neglected, it can be taken into account using the linearized approximation to such perturbations of the field equations. And indeed such linearized perturbations are needed to compute the Peierls bracket between components of the curvature tensor~\citep{dewitt05}.

\subsection{Some Hints from Newtonian Gravity}
\label{newtonianhints}

We can gain further important hints from considering the four-dimensional formulation of Newtonian gravitational theory (see \citep{stachel06}, which has references to earlier work). In this formulation, the chronometry (the unique absolute time foliation) and geometry (the preferred class of spatial frames of reference, linearly accelerated w.r.t. each other, and each having a Euclidean three-metric) are fixed, independent non-dynamical structures; this allows one to form non-dynamical four-volume elements (see Section~\ref{ucprmeasurability}) that, as we have seen, are necessary to calculate the space--time averages of any dynamical quantities. The only dynamical structure in the theory is an affine connection, representing the inertio-gravitational field, which is required to be compatible with both the chronometry and geometry.

 It turns out that these restrictions suffice to make the connection trace-free, so that it is actually a \emph{projective connection}. The components of the projective curvature tensor, formed from this connection, are the gauge-invariant, physically significant quantities, measurable in principle (see Section~\ref{ucpr}). Just as in general relativity, the field equations of this version of Newtonian theory relate the Ricci tensor, formed from the projective curvature, to the stress--energy--momentum tensor, formed from masses, velocities and stress tensors of the material sources of the field. And since the four-volume structure is based on the fixed chronometry and geometry, and thus quite independent of the dynamical projective connection field, one can take averages of these components over the four volume and proceed to consider problems of their quantization without having to worry about quantization of the four-volume. 

Minkowski space--time may be regarded as that special solution to the field equations of general relativity, for which the inertia-gravitational field is flat, i.e., for which the affine curvature tensor vanishes, and for which the symmetry group is the ten parameter inhomogeneous Lorentz group. This consists of the six-parameter homogeneous Lorentz group, including spatial rotations and space--time pseudo-rotations (Lorentz transformations), plus the four-parameter inhomogenous part, consisting of spatial and temporal translations. .

One may ask: what is the analogue of Minkowski space--time for Newtonian space--times? It is the special solution to the Newtonian gravitational field equations that is affine-flat, i.e., for which the Newtonian projective curvature tensor vanishes. This is called Galilean space--time, with a Galilean geometry (see~\citep{yaglom}). Its symmetry group includes the six-parameter homogeneous Galilei group, consisting of spatial rotations and space--time shears (Galilei transformations), plus the four-parameter inhomogenous part, consisting of spatial and temporal translations. The big difference from the case of Minkowski space and its inhomogeneous Lorentz group is that, when we ``turn on" the Newtonian gravitational field, the chronometry and geometry do not change, although their symmetry group becomes larger. Only the projective connection changes, leading to a non-vanishing projective curvature tensor, which changes the symmetry group of the geometry from the ten-parameter inhomogeneous Galilei group to a much larger diffeomorphism group, with a $G$-structure that preserves the homogeneous Galilei group locally (see Section~\ref{geometries}).

There is another interesting point about this formulation. Contrary to the usual formulations of Newtonian theory, in which only an electric-type gravitational (electro-gravitational) field occurs, dependent on the positions of the sources relative to the chosen frame of reference, in this formulation, just as in electromagnetic theory, the total field also includes a magnetic type gravitational (magneto-gravitational) field, dependent on the angular velocity of the sources. In other, hopefully more familiar, words, just as in general relativity, a rotating Newtonian mass will drag inertial frames with it. So the recent experimental verification of this order $(v/c)$ effect, in appropriate units, which has been hailed as a crucial experimental test of general relativity~\citep{phillips}, can actually be explained at the Newtonian level. 

\subsection{Near fields, far fields and matched asymptotic expansions}
\label{matchedasymptotic}

As discussed in the previous section, there are two important limiting cases of general relativity. The well-known Minkowski space of SR is the starting point for the weak field, fast motion approximation procedure. 

The other, Newtonian gravitational theory, is the starting point for the strong field, slow-motion approximation procedure, i.e., an approximation in successive powers of $(v/c)$ to solutions of the full Einstein theory. Gravitational radiation does not enter the picture until order $(v/c)^5$, so any quantization effects at lower orders of the near gravitational field tied to the sources are basically due to the effects of quantization of these sources (for quantization of the Newtonian field, see~\citep{christian}). In this approach, free gravitational radiation fields and their quantization (``gravitons" in the popular parlance) only start to enter the picture at order $(v/c)^5$. This is related to the circumstance that, in contrast to electromagnetism, there is no dipole gravitational radiation. Gravitational radiation only starts to be produced by the varying quadrupole (or a higher) moment of its source.

But there is a much better way to handle gravitational radiation than sticking to such a high order of the slow motion approximation. One can restrict the strong field, slow motion, approximation to the near field, starting from Galilean space--time, and use the weak field, fast motion expansion for the far field. Then one can use the method of matched asymptotic approximations, first applied to gravitation theory by William Burke~\citep{burke}, to match the two approximations in the intermediate region. Kip Thorne, Burke's thesis advisor, explained this approach:

\begin{quotation}
Previous work on gravitational-wave theory has not distinguished the local wave zone from the distant wave zone. I think it is useful to make this distinction, and to split the theory of gravitational waves into two corresponding parts: part one deals with the source's generation of the waves, and with their propagation into the local wave zone; thus it deals with ... all of space--time except the distant wave zone. Part two deals with the propagation of the waves from the local wave zone out through the distant wave zone to the observer ... The two parts, wave generation and wave propagation, overlap in the local wave zone; and the two theories can be matched together there. ... [F]or almost all realistic situations, wave propagation theory can do its job admirably well using the elementary formalism of geometric optics.~\citep[p. 316]{thorne}
\end{quotation} 

If one looks at this more carefully, Thorne has actually introduced three zones: 
\begin{enumerate}
\item{The near zone, in which the near field is generated by its source.}
\label{one}
\item{An intermediate zone, in which the transition takes place between zones 1 and 3.} 
\item{The far zone, in which the pure radiation field has broken free from the source.}
\end{enumerate}

Anticipating the discussion below, it is our conjecture that: in the near zone 1, the projective structure dominates because the propagation of the near field takes place along a family of timelike autoparallel curves. In the far zone 3, the conformal structure dominates because the radiation field obeys Huygens' principle and propagates entirely on a family of null hypersurfaces. In the intermediate zone 2, the compatibility conditions between the conformal and projective structures dominate, assuring that the field propagating from zone 1 into zone 2 and the field propagating from zone 2 into zone 3 are actually one and the same field. 

\section{Geometries and their symmetry groups}
\label{geometries}

A couple of important conclusions can be drawn from this discussion. One does not necessarily have to quantize all space--time structures in a theory that incorporates the equivalence principle. And when one does, it is the curvature of some connection that is the prime candidate for quantization. So now let us turn finally to the question posed above: which connection? To answer this question, we must first discuss the broader question: given a physical theory, what space--time structures does it involve?
 
Generally speaking (pun intended), the general theory of relativity is discussed as if it were based on one space--time structure, the (pseudo-Riemannian) metric and one symmetry group, the diffeomorphism group\footnote{Diffeomorphisms are active point transformations, as opposed to passive coordinate transformations.}. Actually, there are many possible \emph{automorphism} or \emph{symmetry groups} for the various space--time structures that may be associated with different physical theories. If one starts from a bare differentiable manifold, together with the collection $L(x)$ of linear frames at each $x$ point of $M$, then the symmetry group of its geometry is the full four-dimensional diffeomorphism group $Diff(M)$, which induces the group of linear transformations $GL(4, \mathbb{R})$ of the frames at each point $x$ of $M$. As one introduces additional space--time structures, the symmetry group will generally be reduced to a subgroup of $Diff(M)$, inducing a corresponding subgroup of the linear frame transformation group:

\begin{quotation}
Let $M$ be a differentiable manifold of dimension $n$ and $L(M)$ the bundle of linear frames over $M$. Then $L(M)$ is the principal fiber bundle over $M$ with group $GL(n;\mathbb{R})$. Let $G$ be a Lie subgroup of $GL(n;\mathbb{R})$. By a $G$-structure on M we shall mean a differentiable subbundle $P$ of $L(M)$ with structure group $G$.~\citep[p.1]{kobayashi} 
\end{quotation}

So the study of possible geometries on a differentiable manifold $M$, and their relation to each other, is equivalent to the study of all $G$-structures on $M$. As we have seen, for example, the symmetry group of Newtonian gravitation theory at each point of space--time is the Galilei group that preserves the four-volume of all frames. The compatibility conditions between affine connection and the chronometry and geometry then require the preservation of these conditions under parallel transport; so the full symmetry group must be a subgroup of $SDiff(M)$, the group of unimodular diffeomorphisms. These conditions place severe restrictions on the affine curvature tensor, but, as discussed above, they leave just enough freedom to introduce Newtonian gravitational fields of both electric and magnetic types. The existence of an invariant four-volume element means that average values of the components of the affine curvature tensor are meaningful concepts, and indeed the \emph{tidal forces} serve as a means for their measurement.

\section{Unimodular conformal and projective relativity (UCPR)}
\label{ucpr}

With this background, let us turn finally from discussion of approximations to the case of the full theory of general relativity. As early as the 1920s, Hermann Weyl discussed~(see \cite{weyl23}, Section 23, and for the development of his views~\cite{scholz}) the two basic conditions needed for general relativity: the existence of 1) straight lines and 2) null elements. As far as it goes, this is correct; the first leads to the projective structure, and the second to the conformal structure. But for our purposes it is important to add a third item. As Weyl also noted \citep[Section 19]{weyl23}, in order to distinguish between similarity and congruence of geometrical figures, it is necessary to introduce 3) an invariant four-volume structure, preferably path independent.

Weyl did not lay equal stress on this third element because his aim at the time was to develop a unified field theory based on a geometry (now called a Weyl space) that violates this condition. And if one puts the conditions on a Weyl space that restrict the conformal geometry to a single metric, one jumps immediately to the pseudo-orthogonal subgroup $SO(3,1)$ of $SL(4,\mathbb{R})$. 

But for our purposes, the initial introduction of $SL(4,\mathbb{R})$ allows the definition of invariant space--time averages of other physical quantities, and hence the investigation of their measurability and co-measurability. As we have seen, this is a crucial preliminary to the formulation of a quantum version of a theory based on such quantities. 

Mathematically it is simple to introduce condition 3) for any geometric structures on a differentiable manifold prior to, and independently of, conditions 1) and 2). The representations of the traditional symmetry group of general relativity, $Diff(M)$ can be neatly split into two parts using the fact that its stabilizer (also called isotropy subgroup) at a point of $M$, $GL^{+}(4, \mathbb{R})$, can be expressed as a group product of two other groups, $GL^{+}(4,\mathbb{R})\cong\mathbb{R}^{+}\times SL(4,\mathbb{R})$. As the Wikipedia entry on the representations of $GL^{+}(4, \mathbb{R})$ states\footnote{While Wikipedia is not always a reliable reference, in this case we have not found a more concise explanation of the physical significance of the break-up of the diffeomorphism group.},
\begin{quotation}
We know the reps of $SL(n,\mathbb{R})$ are simply tensors over $n$ dimensions. How about the $\mathbb{R}^{+}$ part? That corresponds to the density, or in other words, how the tensor rescales under 	the determinant of the Jacobian of the diffeomorphism at $x$. ... So, we have just discovered the tensor reps (with density) of the diffeomorphism group.~\citep{wikipedia}
\end{quotation}

If one starts from $SDiff(M)$, the group of unimodular diffeomorphisms mentioned above, as the automorphism group of any geometry, there is no need to introduce densities; condition 3) is already incorporated in this requirement, and such structures as the projective affine connection and one-form, and conformal metric and scalar field can each be introduced independently, and relations between them formulated later. 

The physical interpretation of these elements and their relations can also be made independently by an approach assuming that the unimodular group is the maximum possible automorphism group of any physical theory~\citep{stachel11,bradonjic}. As stated above, if we start with a bare manifold, the local geometry is determined by the local diffeomorphism group $Diff(M)$. Once the tangent and dual co-tangent fibre bundles are introduced, we can introduce an invariant four-volume structure by reducing the symmetry group to $SDiff(M)$, the unimodular group. This four-volume structure can be represented mathematically by a scalar field that is dual to the representation by the exterior product of the four basis vectors. (It is a scalar field and not a density because $SDiff(M)$ eliminates the distinction between the two).

Furthermore, the unimodular group allows us to independently introduce:
\begin{itemize}
\item{a symmetric traceless projective connection which determines the autoparallel paths on the manifold}
\item{an \emph{affine one-form} (or projective one-form, as it is called in some literature), which determines the affine parameter along the auto-parallel paths determined by the projective connection}
\end{itemize}

The autoparallel structure of a space--time can be identified with the projective connection, which together with a covector field, the affine or projective one-form, uniquely determine a torsion-free linear affine connection. The projective connection defines the autoparallel paths and the projective one-form determines the preferred parametrizations making them autoparallel curves. Under $Diff(M)$, the components of the projective structure do not transform like the components of a linear affine connection. All dynamics of ponderable matter is based on the projective connection and the affine one-form. 

Finally, the causal structure of a space--time is defined by the conformal metric that, together with the scalar field representing the volume form, uniquely determine a pseudo-Riemannian metric (with Lorentz signature). The conformal metric determines the null cone structure in the tangent space at each point, and hence the null geodesic paths; while the scalar field determines their preferred parametrization, and hence the tangent null vectors. Again, under $Diff(M)$, no such invariant breakup is possible.

Let us emphasize two points:
\begin{enumerate}
\item{The scalar and covector fields are independent of the conformal metric and the projective (and of course the conformal) connection. This means that the definition of a four-volume field is independent of all other space--time structures.}

\item{Even more remarkably, given a volume form and projective connection, there always exists a projectively related affine connection that is equiaffine, i.e., such that the volume form is parallel (see~\citep[Theorem 4.3]{sanchez} and \citep[Proposition 4]{belgun}}.
\end{enumerate}

Most significantly for measurability, this means that in UCPR projective one-form and volume scalar can be related in such a way that the parallelism of the volume form holds quite independently of the subsequent choice of a conformal metric or projective connection. Thus, the ability to define four-volume averages of the projective curvature tensor, discussed above for Newtonian theory, now exists in general relativity, and can even be extended to include four-volume averages of the conformal curvature tensor. Contrary to approaches based exclusively on the metric tensor, or even those based on the metric and connection, in UCPR one can separate the question of a possible discrete structure of space--time from the problem of quantizing the other dynamical structures on that manifold.

Of course, this does not mean that we preclude the study of such a possible discrete structure by quantization of the projective one-form and scalar field, but merely that we do not have to confront this problem before taking space--time averages of these other dynamical variables. Indeed, it was the inability to make such a separation that forced~\citep{bergman} to use four-volumes in the background Minkowski space--time for the calculation of such averages of the linearized Riemann tensor components.

It also follows that a parallel volume scalar fixes a unique parametrization (up to linear transformations) of all autoparallel paths, turning them into autoparallel curves~\citep[Theorem 4.3]{sanchez}). When a causal structure is introduced, this means that a proper time is now defined for timelike autoparallels, as well as a proper length for spacelike ones. Of course, the compatibility of the causal and projective structures would assure that causal geodesic paths coincide with projective autoparallel paths; but this does not per se assure agreement on their parametrization~\citep{bradonjic}. In UCPR, parametrization of autoparallels and agreement between geodesics and autoparallels become independent questions.

\section{UCPR and Measurability}
\label{ucprmeasurability}
We shall now discuss the UCPR approach to measurability analysis in quantum gravity, which is based on the distinction between the near (induction) field produced by massive bodies and the far (radiation) field that has escaped these material sources. Two limiting spacetimes result, each forming the starting point for a method of approximating solutions to the general-relativistic field equations. In UCPR, the treatment of both the limiting spacetimes and the successive approximations is based on the unimodular symmetry group.

\subsection{Near Field Slow Motion Limit}
This is Newtonian limit, sketched in Section \ref{newtonianhints}. The compatibility conditions between the inertio-gravitational connection and both the Euclidean three-geometry and absolute time constrain the connection to be trace-free, so it is a projective connection. The four-volume at each point is a four- dimensional parallelopiped, consisting of a unit cube, formed by three spacelike orthonormal basis vector fields $\bf{e}_{a}$, $(a=1,2,3)$ that span the Euclidean three space, and a fourth velocity vector field $\bf{V}$ defining a frame of reference. This vector field obeys the condition $<{\bf{V}},{\bf{d}}T>=1$, where $T$ is the absolute time, assuring that one has a unit four-parallelopiped at each point of space--time. The allowed symmetry transformations of the $\bf{e}_{a}$ produce spatial rotations of the cube; and the allowed transformations of $\bf{V}\longrightarrow\bf{V}^{'}$, (see \citep{yaglom}), are restricted by the condition $<{\bf{V}^{'}},{\bf{d}}T>=1$, which ensures that the resulting change in the shape of the four-parallelopiped does not change its unit four-volume. 

The lowest-order term in the components of the projective curvature tensor w.r.t. the basis vectors constitutes the electro-gravitational field, and the next-higher-order term introduces a magneto-gravitational field. Being calculated from the projective connection, both fields are projectively invariant, and four-volume integrals of these components can be calculated. These average values can be ascertained by measuring the motions of finite test bodies acted on by the tidal forces produced by the curvature tensor, as sketched out in Section~\ref{newtonianhints}.

The time-variation of these fields is ultimately responsible for any gravitational radiation in the far field; the method of matched asymptotic expansions allows one to find out just how the two are related (see Section~\ref{matchedasymptotic}). These near fields are of course generated by the motion of material sources, as described by the gravitational field equations relating the projective Ricci tensor to the Newtonian stress--energy--momentum tensor~\citep{stachel06}. The quantization of its sources will produce quantum effects on these otherwise-classical near fields.

\subsection{Far Field, Fast Motion Limit}
The far field, fast motion approximation starts from Minkowski space--time $\tilde{\eta}_{\mu\nu}$, the source-free solution to the Einstein equations. It is assumed to be the limit of a conformal metric tensor field $\tilde{g}_{\mu\nu}$, which obeys the unimodular condition that its determinant is -1. Since we want to use this approximation to treat far-field radiation, we assume that there are no material sources in any of the higher-order approximations (as noted below, electromagnetic radiation field terms are permissible). In the next order of the conformal metric tensor: $\tilde{g}_{\mu\nu}=\tilde{\eta}_{\mu\nu}+\epsilon \tilde{h}_{\mu\nu}$, the condition that $\tilde{h}_{\mu\nu}$ be traceless assures the preservation of unimodularity. The wave fronts (characteristics) will propagate in accord with Huygens' principle, while the family of orthogonal rays (bicharacteristics) will obey Fermat's principle~\citep{hildebrandt}. Their geometric behavior (null characteristic hypersurfaces, and null geodesic rays propagating along a family of spacelike two-surfaces on a null hypersurface) is governed by the conformal metric, while their parametrization (phases of wave fronts and frequencies along the rays) is governed by the scalar field associated with the conformal metric.

The two degrees of freedom of the free gravitational radiation field are determined by the shear of such a family of null rays. The shear is closely related to certain components of the conformal curvature tensor; it is measurable classically in the radiation zone by studying the changing shape of the shadows cast by these rays when an obstacle is placed in their path~\citep[Section 3]{stachel11}. 

Particle-like behavior of the radiation field will only manifest itself when some device producing interactions dependent on the quantum of action is introduced, and discrete effects result that can then be described as particle-like (see~\citep{stachel09} for a discussion of Bohr's views on the applicability of the photon concept). Any such device must consist of two parts:
\begin{quotation}
[D]etection of the state of a quantum field requires at least two stages. At the first stage, some conserved physical quantity is to be transferred from the quantum field to an intermediary device, a quantum system with but a finite number of degrees of freedom. This intermediary device is not the ultimate instrument upon which the outcome of the state determination is registered, because its observable features are subject to the indeterminacies of ordinary quantum mechanics; on the other hand, it is not a quantum field, because its state vector is not subject to second quantization. At the second stage, the conserved quantity is to be transferred from the intermediary device to a classical instrument, whose readout is classically determinate, in such a manner that the state of the quantum field is minimally altered ... Having distinguished between the quantum field, the intermediary device and the classical instrument, we shall avoid ... Bohr and Rosenfeld's word test body, which sometimes seems to refer to the classical instrument, and sometimes to what we have called the intermediary device.' ~\citep[pp. 1147-1148]{bergman}
\end{quotation}

If the sources of the gravitational radiation are charged, or have non-vanishing higher electromagnetic multipole moments, the combined Einstein-Maxwell equations must be used, and the far-field radiation will include both electromagnetic and gravitational components. 


\section{Discussion}

So far, we have emphasized that measurability analysis is necessary for establishing the consonance between the physical concepts posited by a theory and a procedure for their ideal measurement, a step without which the theory is not physically meaningful. The idealized measurement procedure need not, and in fact, stands little chance of serving directly as an actual experiment. However, this does not mean that already established methods of experimental measurement cannot be used as models for an idealized measurement, or that an idealized measurement cannot serve as a prototype for a real measurement. For that reason, it is important to consider not only experiments, but also observations. We shall now briefly discuss the difference between the two for the case of gravitational radiation.

The experiments discussed are all based on the concept of prediction; however there are also observations based on the concept of retrodiction. For a prediction, we focus on the fixed result of an initial act of preparation and sum over all possible outcomes of some final act of registration (classically we sum probabilities, quantum mechanically we sum probability amplitudes). For a retrodiction, we focus on the result of some final act of registration, and sum over all possible acts of preparation that could have led to it\footnote{NB: this has nothing to do with reversing the direction of time!}.

It would be more realistic to attempt to analyze the gravitational radiation detected in order to retrodict the behavior of the massive bodies that were its sources. These sources need not --indeed will not -- be terrestrial, but astrophysical or even cosmological, and thus capable of producing gravitational radiation detectable terrestrially. Indeed, the construction of such gravitational wave detectors and preparations for the correlation of their readings is actually under way. 

There are well known electromagnetic analogs of this procedure. Observational astronomy is actually based on retrodiction. That is, from the taking of a (final) measurement (observation) we are trying to retrodict what were the antecedent circumstances in the source that led to this observation. 

Quantum mechanically, this means we must construct a probability amplitude relating a fixed registration result (our observation) to the various possible acts of production that could have led to this result. The ensemble approach to quantum mechanics is quite capable of handling such a problem. Indeed, the fact that the final registered result may be interpreted as a single photon impinging on the registration device by no means implies that the initial electromagnetic preparation interaction involved a single photon, or even that it must be interpreted using the photon picture.

One might object: Any state of the electromagnetic field can be interpreted as a superposition of states in Fock space, each of which corresponds to a definite number of photons with definite energy--momentum, and, \emph{a fortiori}, any preparation results in such a state. The response is: Yes, but the key word here is \emph{superposition}. We have to compute a partial probability amplitude for each of the Fock space states in the superposition, and then add all the partial probability amplitudes to get the total amplitude for the process. One cannot break up the total process into a lot of partial processes by interpreting each of the partial probability amplitudes as a probability for such a partial process. Only the \emph{total amplitude} can be interpreted as a probability for the \emph{total process}. This is what is called quantum entanglement, and it constitutes the entire mystery of quantum mechanics~\citep{stachel97}.

This idea is used in astronomy, even if it is not called ``retrodiction." The quantum explanation of the Hanbury Brown and Twiss effect in terms of photons is an example of such a retrodiction~\citep{brown}. 

Now let us return to the gravitational case. As discussed above, one might attempt to use the weak field, fast motion approximation for the radiation field, and the slow motion, strong field approximation for the motion of the sources (from their varying quadrupole or higher moments) in the near field, using the method of matched asymptotic expansions, only this time for retrodiction. One could discuss the results of using either a classical or a quantum detector for the same detection process. In the case of a quantum detection process, one would expect gravitational effects analogous the electromagnetic Hanbury Brown and Twiss effect.

In the electromagnetic case, the linearity of the Maxwell equations implies that there is no immediate interaction between two radiation states of the field, which means in the particle picture one does not have to sum over photon-photon interactions\footnote{We add ``immediate" because vacuum polarization does produce such higher-order interactions.}. In the gravitational case, due to the non-linear nature of the field equations, there will be such self interactions of the field, which means that in the particle picture one will have to include summation over graviton-graviton interactions. The case of cylindrical gravitational waves~\citep{stachel66}, being the only case for which one can write down the coupled wave equations for functions representing the two states of polarization of the gravitational field, might provide a ``toy model," enabling one to investigate such non-linear interactions at both the classical and quantum levels.

We hope to have demonstrated that our quantum gravity research program has a firm classical foundation, and to have opened several promising avenues for its extension into the quantum realm.

\bibliographystyle{elsarticle-harv}
\bibliography{<your-bib-database>}



\end{document}